\pdfoutput=1

\documentclass[11pt]{article}

\usepackage[final]{acl}
\usepackage{multirow}
\usepackage{times}
\usepackage{latexsym}
\usepackage{listings}
\usepackage{algorithm}
\usepackage{algorithmic}
\usepackage{ulem}
\usepackage{amsmath}
\usepackage{booktabs} 
\usepackage{graphicx} 
\usepackage{hyperref}
\usepackage{epstopdf}
\usepackage[T1]{fontenc}

\usepackage[utf8]{inputenc}

\usepackage{microtype}

\usepackage{inconsolata}

\title{Reverse Chain: A Generic-Rule for LLMs to Master Multi-API Planning}

\author{Yinger Zhang\textsuperscript{*,1}, Hui Cai\textsuperscript{*,†,2} , Xierui Song\textsuperscript{2}, Yicheng Chen\textsuperscript{2}, Rui Sun\textsuperscript{2}, Jing Zheng\textsuperscript{2}\\
Zhejiang University\textsuperscript{1} \\ 
Ant Group\textsuperscript{2}\\
\texttt{zhangyinger@zju.edu.cn} \\
\texttt{\{biyu.ch,songxierui.sxr,yicheng.chen,zhengxi.sr,jing.zheng\}@antgroup.com}
}

\begin{document}
\maketitle

\renewcommand{\thefootnote}{}
\footnotetext{*Equal Contributions}
\footnotetext{†Correspondence to biyu.ch@antgroup.com}

\begin{abstract}
While enabling large language models to implement function calling (known as APIs) can greatly enhance the performance of Large Language Models (LLMs), function calling is still a challenging task due to the complicated relations between different APIs, especially in a context-learning setting without fine-tuning. This paper introduces ``Reverse Chain'', a controllable, target-driven approach designed to empower LLMs with the capability to operate external APIs only via prompts. Recognizing that most LLMs have limited tool-use capabilities, Reverse Chain limits LLMs to executing simple tasks, e.g., API Selection and Argument Completion. Furthermore, to manage a controllable multi-function calling, 
Reverse Chain adopts a generic rule based on a backward reasoning process. This rule determines when to do API selection or Argument completion. To evaluate the multi-tool-use capability of LLMs, we have released a compositional multi-tool task dataset, available at \url{https://anonymous.4open.science/r/reverse-chain-8681}. Extensive numerical experiments validate the remarkable proficiency of Reverse Chain in managing multiple API calls.

\end{abstract}

\section{Introduction} \label{sec1}

\begin{table*}[h] 
\centering
\begin{tabular}{cccc}
\toprule[1pt]
Task Type      & Example                                                                                                                        & API planning                                                                                                                                      \\ \hline
Single-tool & What's the weather in New York ?                                                                                       & \textit{getWearther}(city='New York')                                                                                                                         \\ \hline
Independent multi-tool   & \begin{tabular}[c]{@{}c@{}}What's the weather in New York?\\ When's my next meeting?\end{tabular} & \begin{tabular}[c]{@{}c@{}}\textit{getWearther}(city='New York')\\    \textit{showCalendar}(event='next meeting')\end{tabular}                                           \\ \hline
Compositional multi-tool & \begin{tabular}[c]{@{}c@{}}I'm Lucas, Could you find a flight\\ and book it to my destination ?\end{tabular}       & \begin{tabular}[c]{@{}c@{}}\textit{BookFlight}(flight\_ID=\textit{FindFlight}(destination\\=\textit{GetUserDestination}(userName='Lucas'))\end{tabular} \\ \bottomrule[1pt]
\end{tabular} 
\caption{Different task types, classified by the number of required tools and their dependencies for task execution.}
\label{tasktype}
\end{table*}

Recently, there has been an impressive wave in the progress made in Large Language Models (LLMs), due to their excellent performance in  a variety of tasks \citep{chowdhery2022palm,brown2020language,scao2022bloom,wei2022emergent,bubeck2023sparks}. 
However, LLMs still face difficulties with some specialized tasks due to their fundamental limitation on the information they stored and learned,  which can become outdated and may not be suitable for all applications. A practical solution is to augment LLMs with external tools (known as APIs). In this setup, LLMs act as controllers, not only to understand user intents but crucially to select and orchestrate the appropriate tools to complete tasks.

Unfortunately, LLMs still lack the sophistication to fully understand human instructions and effectively implement function calling. Many works are dedicated to enhancing the function calling abilities of LLMs through fine-tuning or in-context learning methods. \citep{patil2023gorilla,qin2023toolllm,schick2023toolformer,tang2023toolalpaca,parisi2022talm,li2023api,liang2023taskmatrix,song2023restgpt,xu2023tool} Compared to fine-tuning, in-context learning approaches offer a more straightforward and scalable solution, as they eliminate the need to train an entirely new model for each new API. Consequently, the primary goal of this paper is to enhance the API planning capabilities of LLMs within the in-context learning setting.

Different from the aforementioned studies which focus on simpler tasks, such as single-tool task or independent multi-tool task (detailed in Table \ref{tasktype}), this paper targets at enhancing LLMs' ability to handle more complicated \textbf{compositional multi-tool task} (detailed in Table \ref{tasktype}). Implementations of this task requires to employ multiple, potentially interdependent APIs, which is common in real-world scenarios but poses a greater challenge in API planning for LLMs. It's worth noting that single-tool task and independent multi-tool task can be seen as subsets of compositional multi-tool task, and the proposed approach can also manage them with minimal modifications. The generalizability of the proposed method to different task types will be discussed in the Section \ref{conclusion}.

\begin{figure}[h]
  \centering
  \includegraphics[width=1\linewidth]{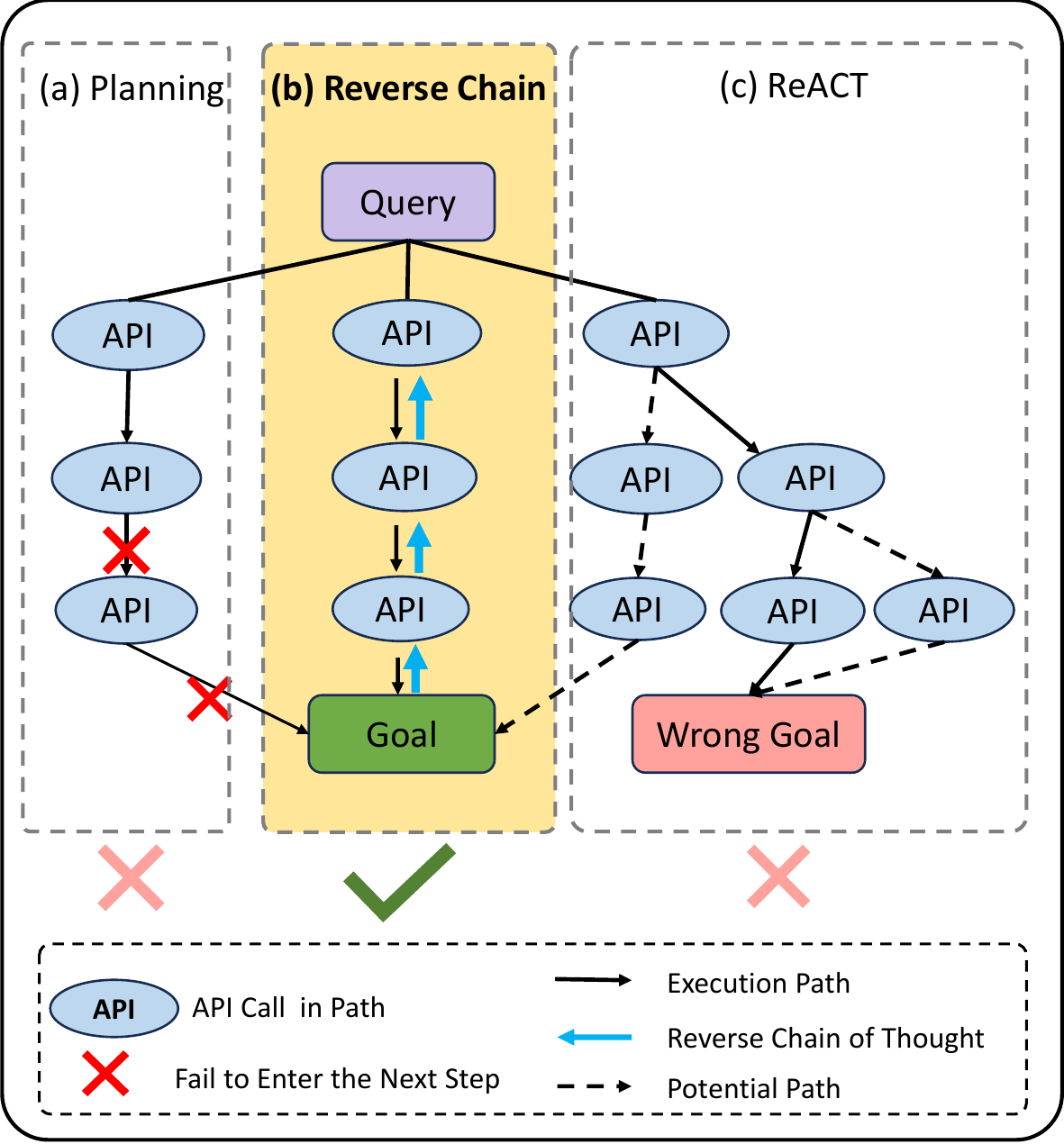}
  \caption{A comparison of our Reverse Chain with the one-step/CoT Planning and ReAct for multi-API planning.}\label{introduction_compareAlgorithms}
\end{figure}

In the realm of tool-use, various prompting techniques have been explored. \textbf{One-step planning} algorithms are introduced in \citep{shen2023hugginggpt, liang2023taskmatrix}, but its accuracy is often low in complex, ambiguous scenarios. The Chain of Thought (CoT) approach \citep{wei2022chain} counters this by step-by-step planning with intermediate reasoning. Known as \textbf{CoT planning}, this technique decomposes tasks into several simpler sub-tasks, thereby boosting reasoning and accuracy. Nevertheless, as illustrated in Figure \ref{introduction_compareAlgorithms} (a), a limitation of these planning methods is their potential for errors in the intermediate stages. While the final step of the plan is intended to achieve the ultimate goal, errors in the intermediate planning steps can lead to execution failures. For instance, as illustrated in the compositional multi-tool case of Table \ref{tasktype}, if the value of `destination' parameter is parsed incorrectly, e.g., destination = `None', it is obvious that \textit{BookFlight} could not be executed successfully. To bridge this gap, \textbf{ReAct}, as described by \citep{yao2022ReAct}, refines reasoning by combining actions and observations for deeper insights. Expanding on this, tool-learning projects \citep{song2023restgpt, ruan2023tptu} utilize the output from each step to inform the next decision. However, as depicted in Figure \ref{introduction_compareAlgorithms}(c), in the multi-function call scenarios, ReAct, despite successfully executing each step, may not adhere to the correct reasoning path towards the final goal, as a result, it deviates to the wrong destination and may end up early. For instance, in the previously mentioned scenario, the ReAct execution flow would be: \textit{GetUserDestination} (userName=`Lucas') -> destination, 
flight\_ID = \textit{FindFlight} (destination) -> Final Answer, which is not completed since the last API \textit{BookFlight} has not been executed.

In summary, both one-step/CoT planning and ReAct encounter significant control challenges: each step of these methods exhibit a high level of \textbf{unpredictability} and \textbf{uncertainty}. Errors can propagate from a wrong thought or action, leading to incorrect solution paths or final goals. This issue arises because these methods start from scratch and progress forward towards the final target, with the LLM bearing the entire burden of planning.

To address these issues, we propose a controllable yet general framework called \textbf{Reverse Chain}. This framework consists of a generic rule and two key modules: API Selection and Argument Completion, both centered on prompting an LLM.  Specifically, the generic rule in Reverse Chain performs a multi-API planning task in a backward manner: it starts by selecting the final API for a task, and then completes the required arguments, drawing values from the query and context, or by outputs of other APIs. When a new API is selected during the argument completion stage, this process repeats. The procedure continues iteratively until all arguments of all APIs are filled.
Reverse Chain distinguishes itself from previous work with the following three main advantages: 1. \textbf{Backward reasoning}, starting from the final goal, preventing planning from deviating into a wrong direction (commonly occurs in ReAct), thus ensuring the correctness of the final goal. 2. The \textbf{step-by-step decomposition} dominated by the rule makes the process controllable, with each stage being forward-executable, effectively avoiding errors common in one-step/CoT planning, such as incorrect intermediate stage. 3. The \textbf{tasks of LLMs are simplified} to just selecting APIs and filling arguments, avoiding complex planning. This strategy effectively utilizes the strengths and capabilities of the existing LLMs without depending on extensive reasoning abilities. 

In summary, the \textbf{contributions} of this paper are: 

\begin{enumerate}
\item This paper presents Reverse Chain, a straightforward framework to improve the API planning capabilities of LLMs in an in-context-learning setting. By employing a  backward reasoning scheme and a step-by-step problem-solving methodology, the process becomes more manageable and controllable.
\item To the best of our knowledge, this paper is the first to focus on API planning for compositional multi-tool task.To assess the capabilities of LLMs in handling such tasks, We collect a high-quality dataset containing 825 APIs and 1550 instances for that task, constructed automatically using GPT-4 \citep{OpenAI2023GPT4TR}. Additionally, an automatic evaluator powered by GPT-4 is also developed for the efficient evaluation purpose.
\item Extensive experiments are conducted to demonstrate the superiority of the Reverse Chain approach in multi-API calling tasks, surpassing the state-of-the-art in-context learning approaches, e.g., CoT and ReAct. 
\end{enumerate}

\section{Related Work}
\textbf{Tool Learning} The discussion of tool usage in LLMs has grown significantly, with models like Toolformer leading the way \citep{schick2023toolformer,nakano2021webgpt}. Current approaches can be divided into two categories. The first category focuses on enhancing the tool-specific capabilities of language models through fine-tuning with specialized datasets \citep{patil2023gorilla,qin2023toolllm,schick2023toolformer,tang2023toolalpaca,parisi2022talm,yang2023gpt4tools,qian2023creator}. The second category directly leverages the capabilities of LLMs, prompting them to interact with various tools, ranging from AI models \citep{shen2023hugginggpt,wu2023visual} to more versatile tool sets \citep{li2023api,liang2023taskmatrix,song2023restgpt,xu2023tool}. Generally, the prompting approach is simpler and more scalable, but it still has a significant gap compared to fine-tuning method, so this work is proposed to enhance the API planning capability of prompting methods.
It is notable that while the previously mentioned studies introduced numerous tool-learning datasets, they primarily encompass relatively simple tasks, focusing on single-tool task or independent multi-tool task. In contrast, this paper targets a more complex task called compositional task, where multiple dependent APIs are needed.

\textbf{Prompting LLMs} Various methods, like CoT \citep{wei2022chain} for task decomposition and ReAct \citep{yao2022ReAct} for melding reasoning with action, enhance general prompting capabilities. Additionally, numerous planning methods are tailored for tool-use. \citep{shen2023hugginggpt,liang2023taskmatrix} start by generating a direct solution outline, followed by selecting and executing relevant APIs. DFSDT \citep{qin2023toolllm} can be seen as an improved version of ReAct, 
enables LLMs to evaluate different reasoning paths and select the most promising one. RestGPT's \citep{song2023restgpt} workflow involves an iterative ``plan and execute'' cycle. Meanwhile, \citep{ruan2023tptu} employs a sequential planning approach, feeding the outcome of each step into the subsequent one. All these works require an LLM to perform either full or step-by-step planning based on the task. However, the Reverse Chain proposed in this work simplifies this by having the LLM focus on just two tasks: API selection and argument completion, thereby greatly simplifying the task complexity. Furthermore, Unlike previous methods that progress from scratch to the final goal, Reverse Chain starts from the end goal and reasons backwards, enhancing controllability.

\section{Reverse Chain: A Multi-API Planning Approach} \label{s2}

\begin{figure*}[h]
  \centering
  \includegraphics[width=0.7 \linewidth]{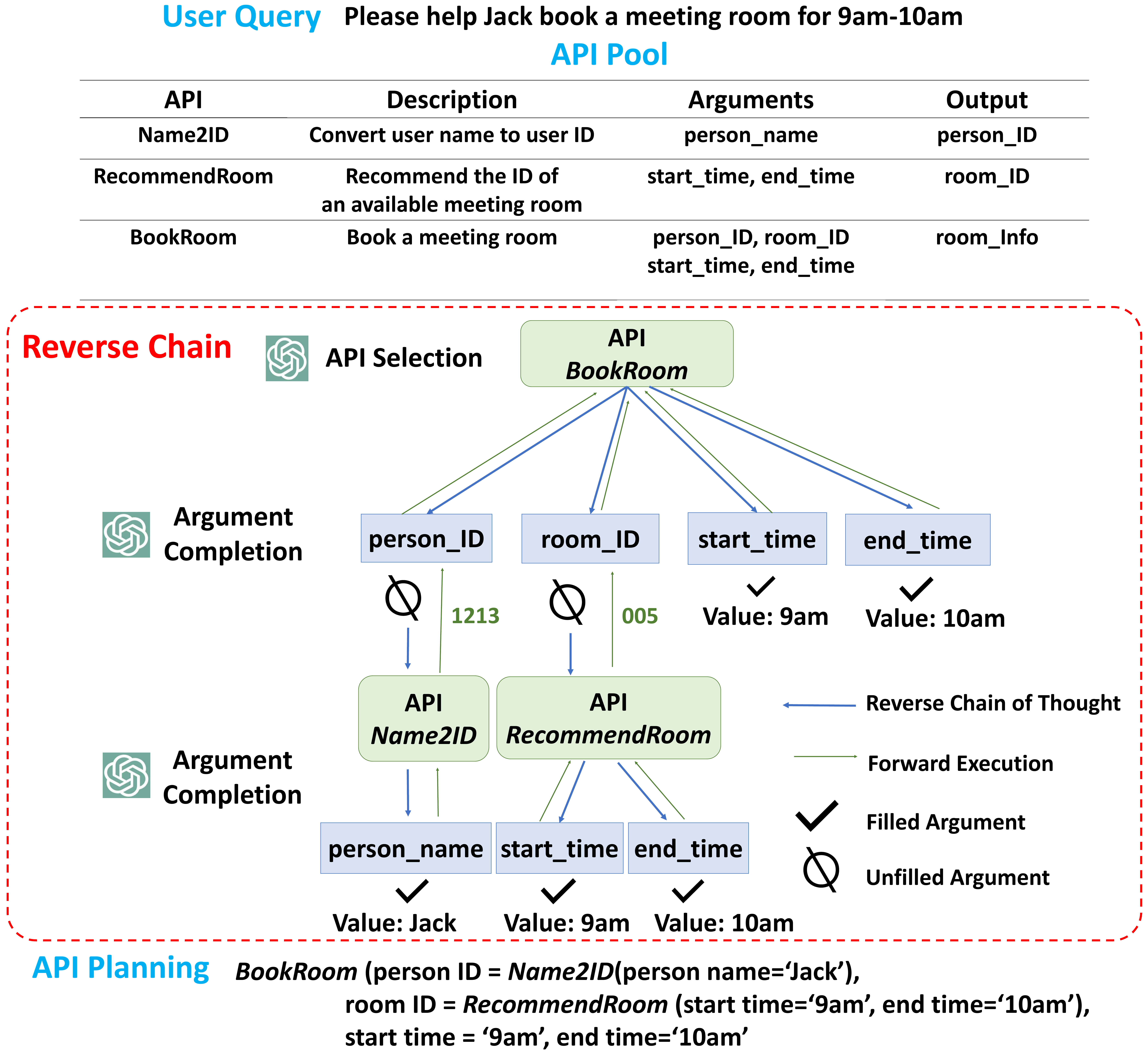}
  \caption{Workflow of Reverse Chain on an example.} \label{reverse model}
\end{figure*}

The objective of this work is to  generate effective API planning based on user queries and API candidates. Figure \ref{reverse model} provides a detailed example: A user query could be a natural language request like ``Please help Jack book a meeting room from 9:00 am to 10:00 am''. Each API in the API pool is characterized by its description, arguments, and output. e.g., the API \textit{RecommendRoom} has a functionality description of ``Recommend the ID of an available meeting room'', arguments ``start\_time'' and ``end\_time'', and an output of ``room\_ID''. A successful API planning consists of two parts: selecting the proper API and filling in all the arguments correctly, where the argument values can come from the query or context, or from the output of another API. 

Section \ref{s21} outlines the Reverse Chain process, while Section \ref{s22}  specifically discusses the two modules that interact with LLM: \textbf{API Selection} and \textbf{Argument Completion}.

\subsection{Reverse Chaining} \label{s21}
Different from CoT and ReAct, Reverse Chain performs a task decomposition in a reverse manner, and its step-by-step problem-solving path is predefined by a generic rule. It is notable that this generic rule is not restricted with a certain type of tasks. 

Figure \ref{reverse model} shows an example of Reverse Chain applied to API planning for a query. Initially, Reverse Chain selects the final API for a given task, this step is referred to as \textbf{API Selection}.  In this example, LLM selects an API named \textit{BookRoom} to match the task ``booking a meeting room''. Next, the required arguments of the selected API are identified through engineering guidance, e.g. API \textit{BookRoom} has four required arguments, that is, person\_ID, room\_ID, start\_time, and end\_time.  There are three possible approaches for arguments filling, and we define this process as \textbf{Argument Completion}: \\
\textbf{Case 1}. The argument value extracted directly from the context and user query, e.g., start\_time = 9:00 am; \\
\textbf{Case 2}. When the argument value could not be obtained directly, Reverse Chain searches for another possible API whose output could complete the missing argument, e.g., the argument person\_ID could be obtained from API \textit{Name2ID}; \\
\textbf{Case 3}. If it is unable to obtain the argument value from the above two cases, the generic rule will request the argument value directly from the user.  

For the selected internal APIs in \textbf{Case 2}, Reverse Chain makes recursive calls to complete the required arguments of these APIs, e.g., the required argument of \textit{Name2ID} is person\_name, and the value `Jack' could be obtained through \textbf{Case 1} in Argument Completion. The algorithm continues until the termination condition is met, i.e., all of the required arguments are completed. Finally, when all required arguments of an API are filled, the API is ready to be executed forward to complete the given task. 

\subsection{LLM Modules in Reverse Chain}\label{s22}
\subsubsection{API Selection}
In this module, the LLM effectively determines the relevant API by analyzing the task descriptions and API candidates. The specific prompt used in this module is depicted in Figure \ref{prompt}.(a). 
Within the Reverse Chain, the API Selection module is employed in two different scenarios, separated with regard to different task description and API candidates. The first scenario occurs when selecting the ultimate API. In this case, the task descriptions correspond to the user query and the API candidates refers to  all APIs in the API Pool.
The second scenario occurs as a sub-module of Argument Completion. When the value of an argument cannot be obtained from the user query or context, the Reverse Chain selects an appropriate API whose output can fulfill the missing argument. In such cases, the task descriptions refers to the description of the unfilled argument. The scope of API candidates can be narrowed down through variable type matching, which encompasses Time, Date, String, etc. This capability facilitates a more refined selection process, leading to a improved accuracy.

\subsubsection{Argument Completion}
After API Selection, the required arguments for the selected API are determined with the help of engineering guidance. In this module, the LLM is leveraged to complete these arguments using information from the query, context and API candidates. The execution follows three possible outcomes:
\\
\textbf{Case 1} 
The argument value is directly extracted from the context or user query. \\
\textbf{Case 2}
Another API is used to complete the missing argument value, indicating that the LLM is unable to obtain the argument value directly. It should be noted that the arguments of this new internal API must be completed before execution. \\
\textbf{Case 3}
 None, indicating the inability to obtain the argument value from the context, user query, and potential API output. In this case, the generic rule will request the argument value directly from the user.

\begin{figure}[h]
  \centering
  \includegraphics[width=1 \linewidth]{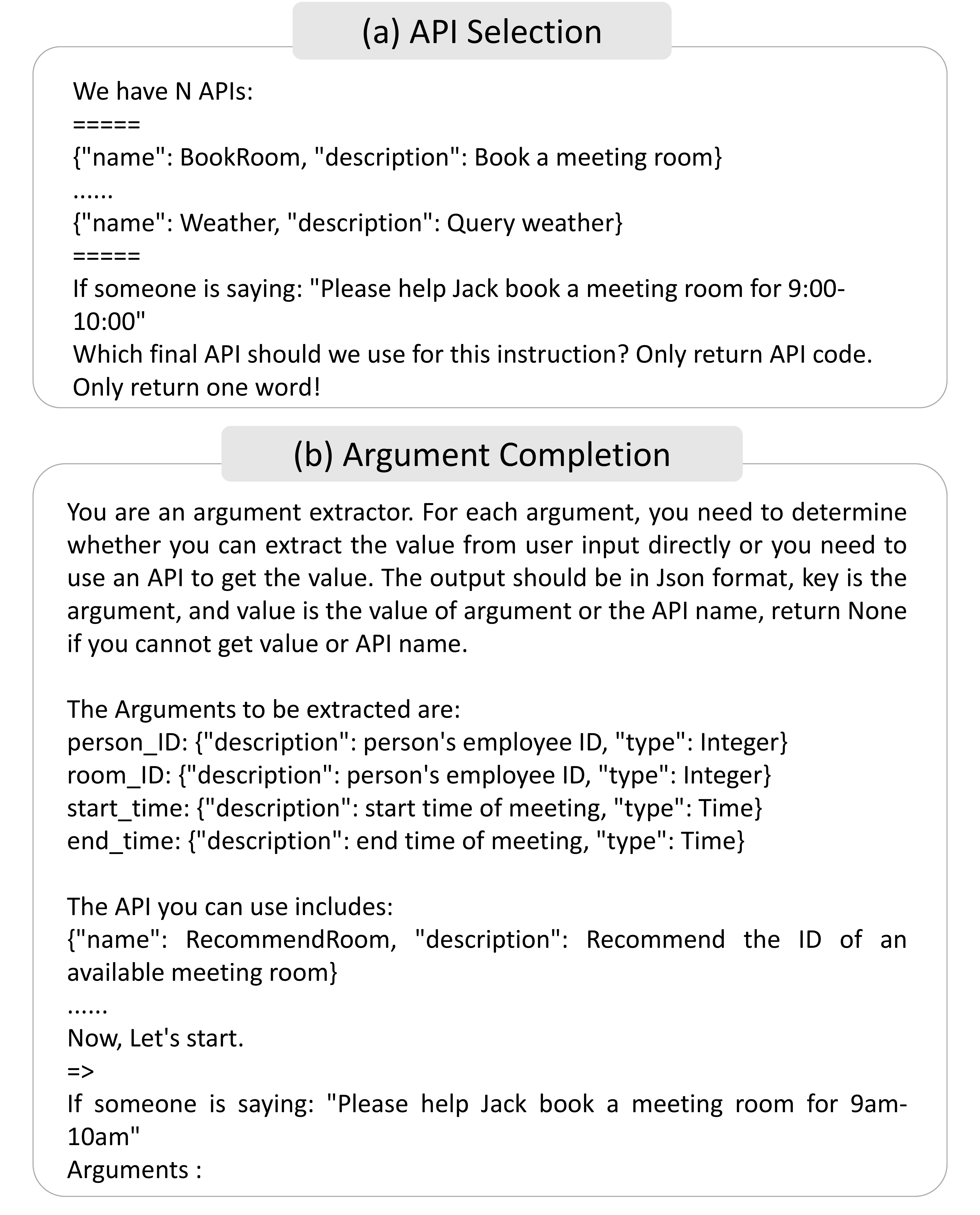}
  \caption{The details of prompts used in Reverse Chain for API Selection and Argument Completion (when LLM is chatgpt).
    }\label{prompt}
\end{figure}

Specific optimizations have been applied to the aforementioned approach, which are further explored in Section \ref{322}. The optimized prompt used in this module is illustrated in Figure \ref{prompt}.(b).

\section{Experiments}
In this section, extensive experiments are conducted to investigate the performance of Reverse Chain. We start with generating an evaluation dataset automatically, benchmarking different in-context learning methods on function calling and defining the evaluation metrics.  In Section \ref{s31}, to benchmark Reverse Chain, we compare its API planning capabilities with the current state-of-the-art in-context learning solutions on ChatGPT. Section \ref{s32}, details a set of ablation experiments designed to elucidate the underlying principles of Reverse Chain. Finally, Section \ref{s33} analyzes the factors contributing to the effectiveness of Reverse Chain.

\textbf{Dataset}
We construct a dataset for evaluating compositional multi-tool tasks. Guided by the self-instruc paradigm \citep{wang2022self}, this dataset is generated automatically based on GPT-4 and ChatGPT (gpt-3.5-turbo), involving the following steps:

\begin{enumerate}
    \item Initially, APIs are selected from public repositories, including API-Bank \cite{li2023api} and public-apis\footnote{https://github.com/public-apis/public-apis}. We then manually create 20 diverse seed examples for compositional multi-tool task, each comprising three components: \{API and its description, User query, System response\}. A specific seed example is detailed in Figure \ref{seed-example} in Appendix \ref{A-dataset} .
    
    \item These seed instances serve as in-context examples for GPT-4, so as to generating more complex new samples. The prompts for GPT-4 are detailed in Figure \ref{sample generation prompt} in Appendix \ref{A-dataset construction}, include a general description of the task, a randomly chosen seed example, and a prescribed response format. Then we conduct manual quality checks to filter out erroneous samples, achieving a 50\% filtration rate. The high-quality samples produced are used as new seed examples for further data collection, repeating the process multiple times. To enhance dataset diversity, GPT-4's temperature is set at 0.8.
    
    \item Additionally, we employ ChatGPT to enhance API information and uniformly standardize the samples into a JSON format. A detailed example is in the Figure \ref{sample} in Appendix \ref{A-dataset}. Each sample includes fields: \{APIs, Query, Label\}, with each API in APIs represented as a JSON object with fields: \{name, description, arguments, output, format\}. Notably, the fields \{arguments, output, and format\} are generated by leveraging existing information. The prompt for this is outlined in Figure \ref{format conversion prompt} Appendix \ref{A-dataset construction}.
\end{enumerate}

It's worth mentioning that the dataset comprises 825 unique APIs across 20 categories, totaling 1550 labeled instances, with the categories detailed in Table \ref{API doamin} in Appendix \ref{A-dataset}. Focused on compositional multi-tool tasks, the samples are classified into three levels based on API nesting complexity: Level-1, two levels of API nesting, containing 798 instances; Level-2, three levels of API nesting, containing 693 instances; and Level-3, more than four levels of API nesting, containing 59 instances. Each Instance has an average of 2.93 function calls.

It is clear that this synthetic dataset is suitable for evaluation since: 1. Automated data generation guarantees unbiased data; 2. The APIs are spread across diverse domains, accurately reflecting real-world situations; 3. The inclusion of various nesting levels in compositional multi-tool tasks ensures a rich diversity.

\textbf{Baseline}
To benchmark Reverse Chain, we measure its performance against five other in-context learning methods: \textbf{Zero-Shot}, \textbf{Few-Shot}, \textbf{Zero-Shot-CoT}, \textbf{Few-Shot-CoT}, and \textbf{ReAct}, using ChatGPT as the underlying LLM. Each method integrates API data into the prompt, utilizing the LLM's in-context learning for API planning. The Zero-Shot approach uses API information and user queries in the prompt, Few-Shot adds extra examples to prompt. Zero-Shot-CoT includes step-by-step instructions, while Few-Shot-CoT adds explanations to these steps in the examples. ReAct, implemented via the langchain framework, uses a (thought, action, observation) format for task execution.
Examples of prompts for these methods can be found in the Appendix \ref{Prompts for baseline method}. Experiments are conducted on two LLMs: GPT-3.5-turbo at the gpt-3.5-turbo-0301 checkpoint with the temperature set to 0.1. 

\textbf{Metrics}
We use accuracy as a metric to evaluate API planning, which consists of two aspects: API name and API arguments. The value of argument consists of direct value filling or another API calling. 

Given the diversity of output formats across solutions, we rule out simple string matching due to its inefficiency and manual annotation for its time-consuming nature. Instead, we craft an efficient automated evaluator using GPT-4. Tailored prompts are designed for each baseline method to match its output characteristics. The prompts are presented in Appendix \ref{evaluation promot}. We manually test 200 samples, comparing human annotations with GPT-4 evaluations, and discover that the GPT-4 evaluator exhibits a strong 89\% correlation with human assessments.

\subsection{Main Results}\label{s31}
Throughout the experiments, the given API candidates set in prompt only includes the needed APIs for a given task since the focus of this paper is primarily on evaluating the capability of LLMs on generating a proper API calling rather than the retrieval of API. Table \ref{main result} compares the accuracy of different in-context learning methods. Under a \textbf{Zero-Shot} setting, the LLM's API planning accuracy stands at approximately 68.97\%. Although \textbf{Few-Shot} methods raises this to 81.87\%, the addition of Chains of Thought (CoT) further elevates performance to 87.16\% in \textbf{Few-Shot-CoT}, which indicates the benefit of decomposing complex tasks. The \textbf{ReAct} strategy, with its reasoning-action-observation approach, also improves upon the zero-shot method. However, the standout performer is the \textbf{Reverse Chain} method, which surpasses all others by simplifying the multi-API calling problem into two easier tasks (API Selection and Argument Completion) and adopting a target-driven approach, thereby minimizing uncertainty. Impressively, \textbf{Reverse Chain} achieves superior results even in a zero-shot context surpassing both the \textbf{Few-Shot-CoT} and \textbf{Few-Shot} methods. Additionally, Table \ref{main result} displays results across different levels of API planning where higher levels indicates greater difficulty. As expected, all methods exhibit increased error rates as the complexity of API planning escalates. In these more challenging scenarios, the Reverse Chain approach demonstrates a more pronounced improvement compared to other methods. This significant gap underscores its robustness and effectiveness in handling complex multi-API calling tasks.

\begin{table}[] \small
\centering
\begin{tabular}{c|cccc}
\toprule[1pt]
Method        & level 1          & level 2          & level 3          & Overall         \\ \hline
Zero-Shot     & 72.06          & 67.68          & 42.37          & 68.97         \\
Few-Shot      & 86.46          & 77.48          & 71.18          & 81.87          \\
Zero-Shot-CoT & 82.45          & 81.38          & 57.62          & 81.29          \\
Few-Shot-CoT  & 89.72          & 85.71          & 66.10          & 87.16          \\
ReAct         & 72.68          & 69.11          & 45.76           & 70.06         \\
\textbf{Reverse Chain} & \textbf{93.99} & \textbf{90.33} & \textbf{86.44} & \textbf{92.06} \\ \bottomrule[1pt]
\end{tabular}
\caption{Evaluation results on various in-context learning methods. We can observe that the proposed Reverse Chain outperforms all other approaches.}
\label{main result}
\end{table}

\subsection{Ablation Study}\label{s32}
In this section, we mainly focus on exploring the impact of creativity of LLMs and different argument completion strategies on the performance of Reverse Chain. The experiments are conducted on GPT-3.5-turbo.

\subsubsection{Creativity and imagination of LLMs on Reverse Chain}
We first investigate the impact of LLM's temperature on Reverse Chain. Temperature controls the randomness of the LLM's output. A lower temperature results in more focused and deterministic responses, while a higher temperature generates more diverse and creative answers. Table \ref{temperature ablation study} shows that Reverse Chain performs better at lower temperatures, with accuracy decreasing when it seeks more creative responses. It makes sense as we require LLM to make rational and accurate decisions.

\begin{table}[h]
\centering
\begin{tabular}{c|cccc}
\toprule
Method          & level 1        & level 2          & level 3          & Overall         \\ \hline
T=0.1 & \textbf{93.99} & \textbf{90.33} & \textbf{86.44} & \textbf{92.06} \\
T=0.5 & 78.45           & 59.88           & 59.32          & 69.42          \\
T=1   & 69.80           & 50.50          & 49.15           & 60.39          \\ \bottomrule
\end{tabular}
\caption {The impact of different temperatures of LLMs on the performance of Reverse Chain. T represents the temperature of ChatGPT}
\label{temperature ablation study}
\end{table}

\subsubsection{Argument Completion Optimization} \label{322}

\begin{table}[h]
\centering
\begin{tabular}{cc}
\hline
Reverse Chain             & \textbf{92.06}            \\ \hline
Reverse Chain\_one-by-one & 74.19                     \\ \hline
Reverse Chain\_three-step & \multicolumn{1}{l}{38.71}  \\ \hline
\end{tabular}
\caption{Ablation study for the design of Argument Completion in Reverse Chain.}
\label{ablation study}
\end{table}

In this part, a series of ablation studies are performed to examine various optimizations during the development of the Reverse Chain Algorithm. The optimizations discussed there primarily concentrate on the stage Argument Completion.

\paragraph {Reverse Chain\_one-by-one} \label{onebyone}
In the existing Reverse Chain method, LLMs simultaneously extracts all argument results. An alternative strategy involves processing each argument completion sequentially, a method we term Reverse Chain\_one-by-one. For instance, the API \textit{FlightBooking} has two arguments: departure\_point and destination. While the standard Reverse Chain completes both departure\_point and destination arguments concurrently, Reverse Chain\_one-by-one first fills the argument departure\_point, followed by the destination.

Table \ref{ablation study} shows that Reverse Chain achieves a 92.06\% accuracy, surpassing Reverse Chain\_one-by-one's 74.19\%. The performance disparity arises because the LLM in Reverse Chain can access to all information about unfilled arguments during the argument completion process. This comprehensive insight enables more precise and accurate argument filling. Consider the API example \textit{FlightBooking} with the user query: ``help me book a flight from London to Los Angeles'', Table \ref{ablation1} demonstrates that in Reverse Chain\_one-by-one, both arguments mistakenly extract the value `London', as the LLM interprets the query's location as the destination. Conversely, Reverse Chain, recognizing two separate arguments for departure\_point and destination, accurately distinguishes between the two locations in the query.

In addition to its superior performance, Reverse Chain is also more efficient in terms of time and computational resources since it only requires one interaction with the LLM.

\begin{table}[h]
\centering
\begin{tabular}{c|cc}
\toprule[1pt]
                          & departure & destination        \\ \hline
One-by-one & London           & \textbf{London (wrong)} \\
Reverse Chain             & London           & Los Angeles        \\ \bottomrule[1pt]
\end{tabular}
\caption{Examples of Reverse Chain\_one-by-one and Reverse Chain}
\label{ablation1}
\end{table}

\paragraph{Reverse Chain\_three-step} \label{three-step}
Here is an example: user query is ``help Jack book a meeting room'', requiring the filling of the person\_ID argument for the API \textit{BookRoom}. In the Argument Completion step of standard Reverse Chain, both the query and API candidate sets are available to the LLM, enabling direct value extraction from the query or API selection. However, in the {Reverse Chain\_three-step} setting, argument completion is further split into two steps: initially, the LLM is given only the query for value extraction, potentially returning the extracted value or `None'. If `None' is returned, then it will move to API selection, choosing from the API candidate set.

Table \ref{ablation study} reveals that Reverse Chain\_three-step attains just a 38.71\% accuracy rate. This is mainly due to the absence of API information during the value extraction step, often leading to forced extraction of incorrect values even when certainty is low. In the given example, the LLM mistakenly identifies `Jack' as the person\_ID value. This confusion is not surprising given the vague nature of the person\_id concept. However, with API information, the LLM can discern between using APIs or forcibly extracting values, thus enhancing accuracy. For instance, the LLM might find that person\_ID is retrievable through the API PersonName2ID, and consequently, it disregards the erroneously extracted `Jack'.

\subsection{Why Reverse Chain works?}\label{s33}

\begin{table}[] \small

\begin{tabular}{c|cc}
\hline
\textbf{}     & \textbf{Wrong Final Tool} & \textbf{Wrong Argument} \\ \hline
Zero-Shot     & 33                        & 132                     \\
Few-Shot      & 29                        & 75                      \\
Zero-Shot-CoT & 36                        & 68                      \\
Few-Shot-CoT  & 22                        & 58                      \\
ReAct         & 91                        & 70                      \\
Reverse Chain &\textbf{ 20}                        & \textbf{40}                      \\ \hline
\end{tabular}
\caption{Error cause statistics all methods.}
\label{error}
\vspace{-2.0em}
\end{table}
In this section, we dissect common errors in API planning and illustrate how the Reverse Chain method mitigates them for improved results. We categorize the errors, identify through manual review, into two primary types, \textbf{Wrong Final Tool} and \textbf{Wrong Argument}, detailed in Table \ref{error}. This statistics is done on 500 randomly sampled instances.

\textbf{Wrong Final Tool} arises when the final API is missing, leading to incorrect API termination and incomplete instructions. This error is prevalent across all comparison methods due to their tendency to plan from the scratch, increases the likelihood of deviating from the final goal. Particularly, ReAct is more susceptible to this mistake because of its thought-action-observation approach that lacks global planning. Reverse Chain, by planning based on the final goal, minimizes this error, except when the query's ultimate intention is ambiguous.

The second error, \textbf{Wrong Argument}, predominates in planning methods, can be further categorized into \textbf{Wrong Argument\_API} and \textbf{Wrong Argument\_Value}. Wrong Argument\_API error occurs when a required argument is the output of another API, but the predicted result bypasses this API, filling in an incorrect value. For instance, the correct argument is person\_ID = \textit{PersonName2ID} (name=`Jack'), but the prediction inaccurately inputs person\_ID=`Jack'. This error often results from mistakes in the intermediate planning steps. In Reverse Chain's argument completion phase, using the optimization approach from Section \ref{three-step}, these errors can be greatly reduced, which allows the LLM to choose between using the API or extracting the argument value. \textbf{Wrong Argument\_Value} involves extracting incorrect values for the argument. Specific cases and optimization strategies for Reverse Chain are discussed discussed in Section \ref{onebyone}.

\section{Conclusion} \label{conclusion}
This paper proposed Reverse Chain, a concise, target-driven approach developed to empower LLMs with the capability to interact with external APIs in an in-context learning setting. By implementing a backward reasoning strategy and a generic rule, Reverse Chain effectively broke down complex function-calling challenges into two fundamental tasks for LLMs: API selection and argument completion. Additionally, we collected a compositional multi-tool dataset for evaluation. Extensive experiments revealed that Reverse Chain markably enhances the tool-use proficiency of the existing LLM ChatGPT, achieving superior performance compared to methods like CoT and ReAct.

Although the current work concentrates on compositional multi-tool tasks, it can also be easily extended to other types of tasks. For instance, in the case of independent multi-tool tasks, after identifying sub-intents at the beginning of the task (known as Intent Detection, a well-established problem in NLP with numerous robust solutions), we could employ the reverse chain process for each identified sub-task separately.

\clearpage
\section{Limitations} \label{discussion}
We identify some limitations with our current work that can be addressed in future work.
\begin{itemize}
    \item The in-context learning approach generally struggles with handling a large number of API candidates due to length limitations. A solution similar to the one in \citep{qin2023toolllm}, which involves adding a retrieval module at the beginning of the pipeline, can be adopted. 
    \item While our demonstration shows that Reverse Chain surpasses other in-context learning methods in performance, it does require more calls to the LLM. This highlights a trade-off between performance enhancement and increased computational resource use.
\end{itemize}

\bibliography{latex/acl_latex}

\clearpage
\appendix

\section{Appendix}
\label{sec:appendix}

\subsection{Sample in dataset}\label{A-dataset}
\begin{figure*}[h]
  \centering
  \includegraphics[width=1 \linewidth]{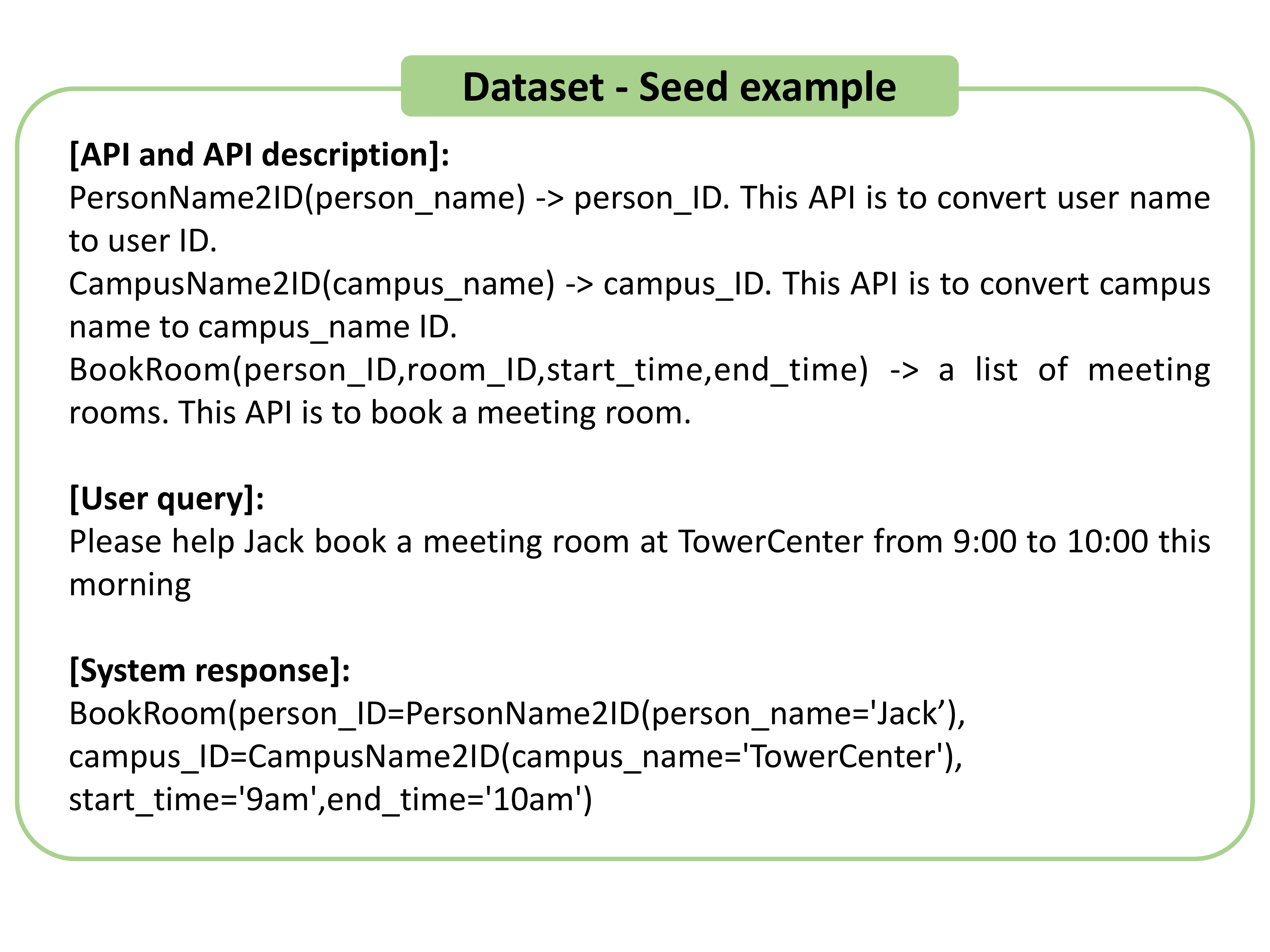}
  \caption{An example of seed example.} \label{seed-example}
\end{figure*}

\begin{figure*}[h]
  \centering
  \includegraphics[width=1 \linewidth]{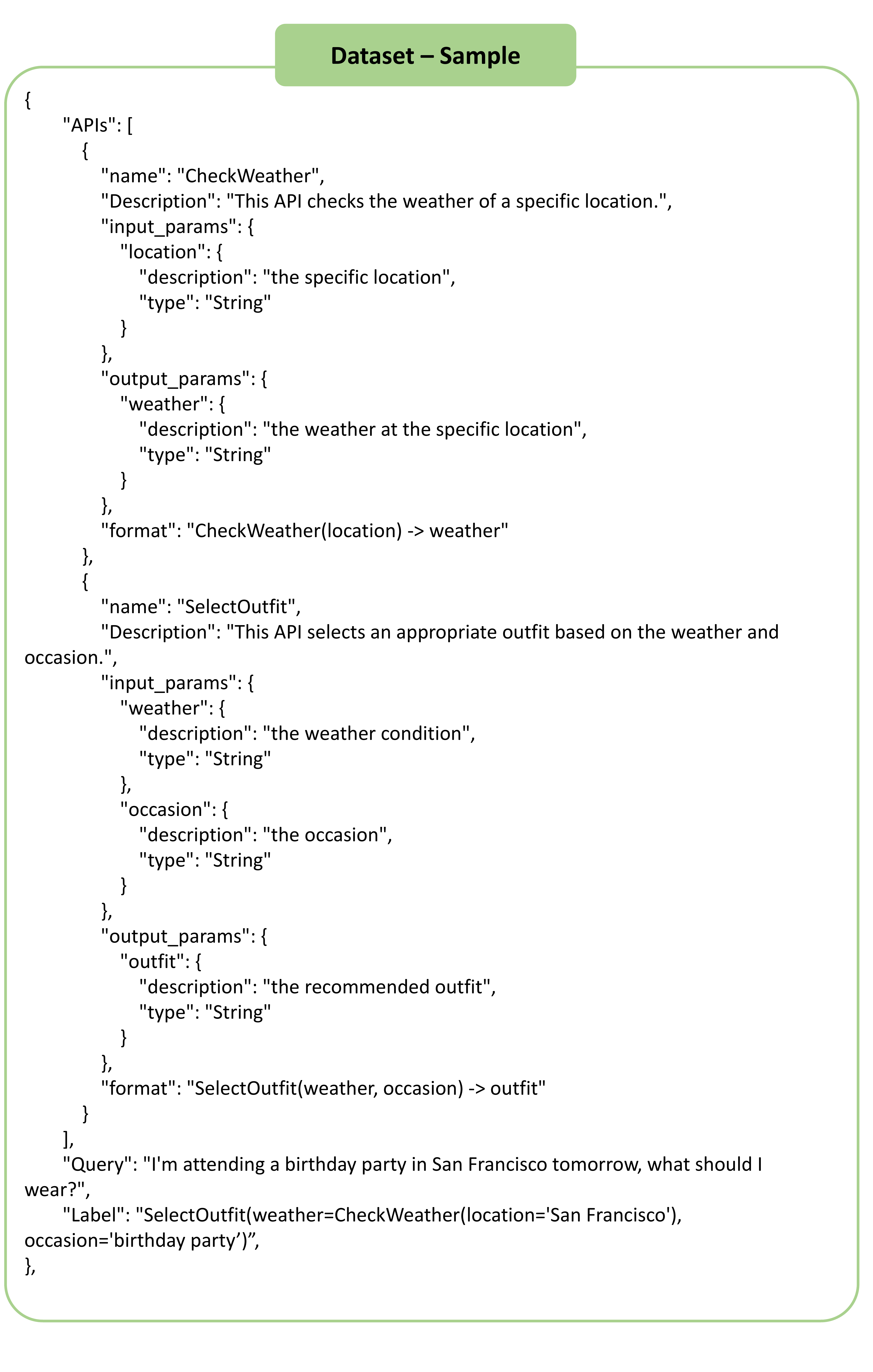}
  \caption{An example of sample in dataset.} \label{sample}
\end{figure*}

In this section, we show the details of the dataset. Figure \ref{seed-example} is an example among the 20 diverse seed examples designed by human. Figure \ref{sample} is an example in the dataset of final version. The category and examples of APIs are listed Table \ref{API doamin}. 

\begin{table*}[]
\begin{tabular}{cc}
\hline
\textbf{Category} & \textbf{example APIs}                                        \\ \hline
Geocoding         & GetDirections,GetUserDietaryRestrictions, DistanceCalculator \\
Weather           & GPS2Weather,WeatherVerification                              \\
Book              & AddBookToReadingList,BooksByAuthor                           \\
Transportation    & FlightBooking,FindFlightByDestination                        \\
Music             & AddSongToPlaylist,MusicConcert                               \\
Food \& Drink     & SearchRestaurant,TableReservation,RestaurantReviews          \\
Entertainment     & CinemaShowtimes,MovieReview, TheatrePlay                     \\
Shopping          & FindProductId,NearestStore, ComparePrices                    \\
Health            & GetExerciseRoutine,NearbyHospitalQuery,GetHealthInformation  \\
Travel            & SearchHotel,CheckBaggageAllowance,PlanTrip                   \\
Database          & CheckInventory,DateConversion                                \\
Calculator        & TaxCostCalculator,CalculateCalorie                           \\
Email             & UserEmail2UserId,SendReview                                  \\
Finance           & InvestmentSuggestion,CountryTaxRate,                         \\
Convertor         & User2Age,HotelName2ID                                        \\
Clothes           & SelectOutfit,OutfitSuggestion,FindClothingType               \\
Time              & ConvertTime,GetEventCalendar                                 \\
Activity          & ActivityBook,PlanDayOut                                      \\
Currency Exchange & CurrencyConversion,GetExchangeRate                           \\
Search            & GetCurrentFuelPrice,ProductSearch                            \\ \hline
\end{tabular}
\caption{Domain distribution and examples of APIs in our dataset.}
\label{API doamin}
\end{table*}

\subsection{Prompts for dataset construction}\label{A-dataset construction}
\begin{figure*}[h]
  \centering
  \includegraphics[width=1 \linewidth]{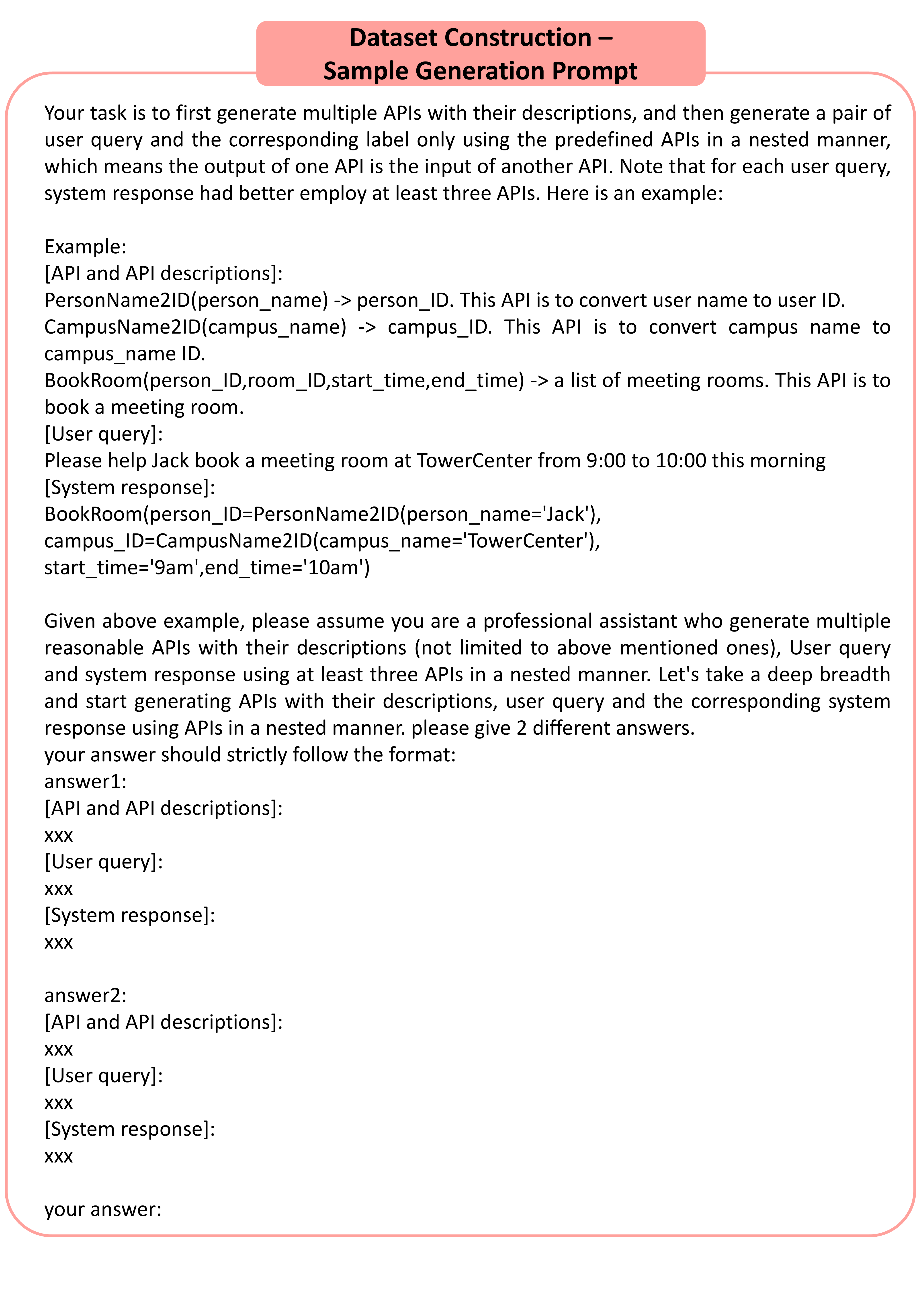}
  \caption{Prompt for new sample generation.} \label{sample generation prompt}
\end{figure*}

\begin{figure*}[h]
  \centering
  \includegraphics[width=1 \linewidth]{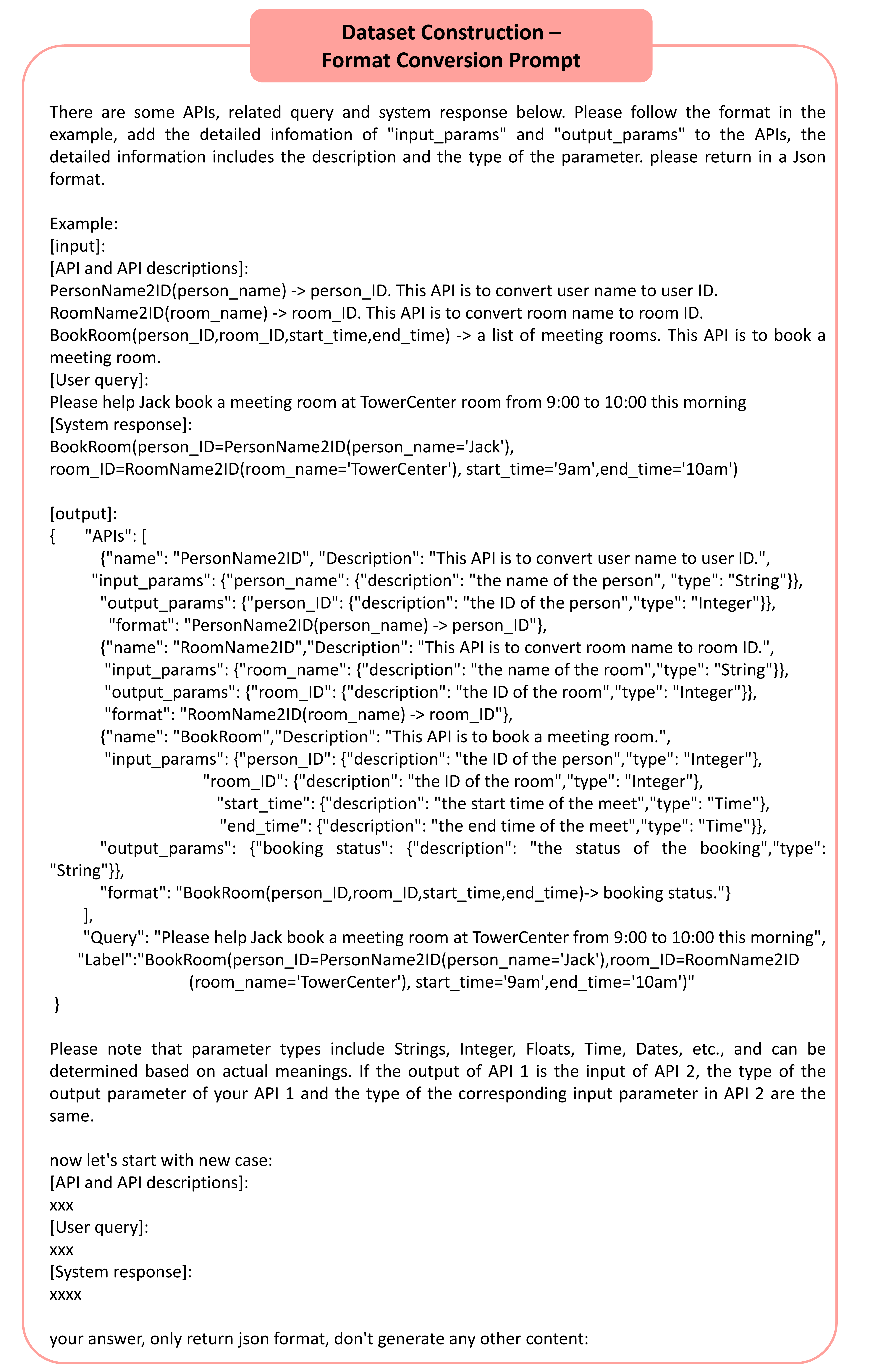}
  \caption{Prompt for Json format conversion.} \label{format conversion prompt}
\end{figure*}

In this section, we show the details of prompt templates in data construction. Figure \ref{sample generation prompt}
is the prompt of new sample generation for GPT-4. Figure \ref{format conversion prompt}
 is the prompt of format conversion for ChatGPT. 

\subsection{Prompts for baseline methods}\label{Prompts for baseline method}
\begin{figure*}[h]
  \centering
  \includegraphics[width=1 \linewidth]{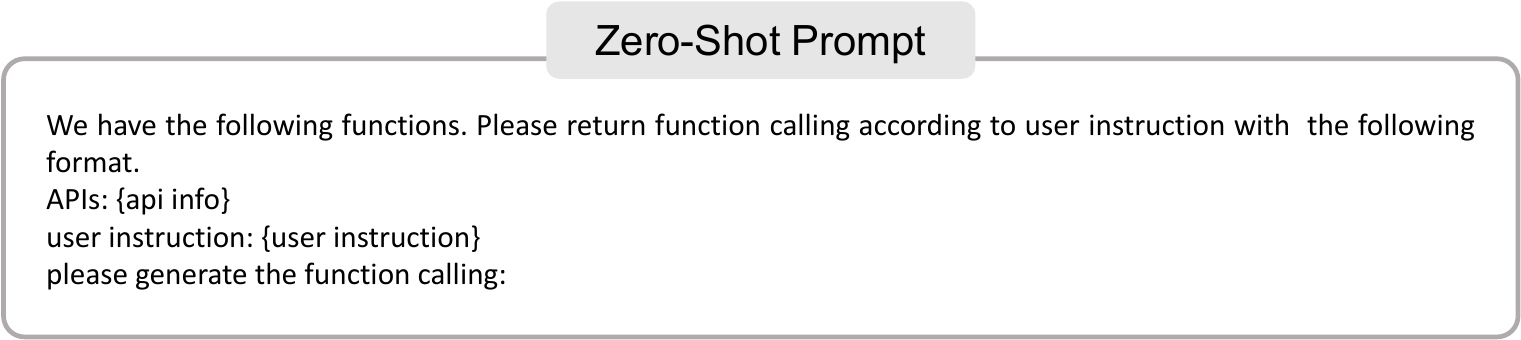}
  \caption{Prompt for Zero-Shot method.} \label{Zero-Shot}
\end{figure*}

\begin{figure*}[h]
  \centering
  \includegraphics[width=1 \linewidth]{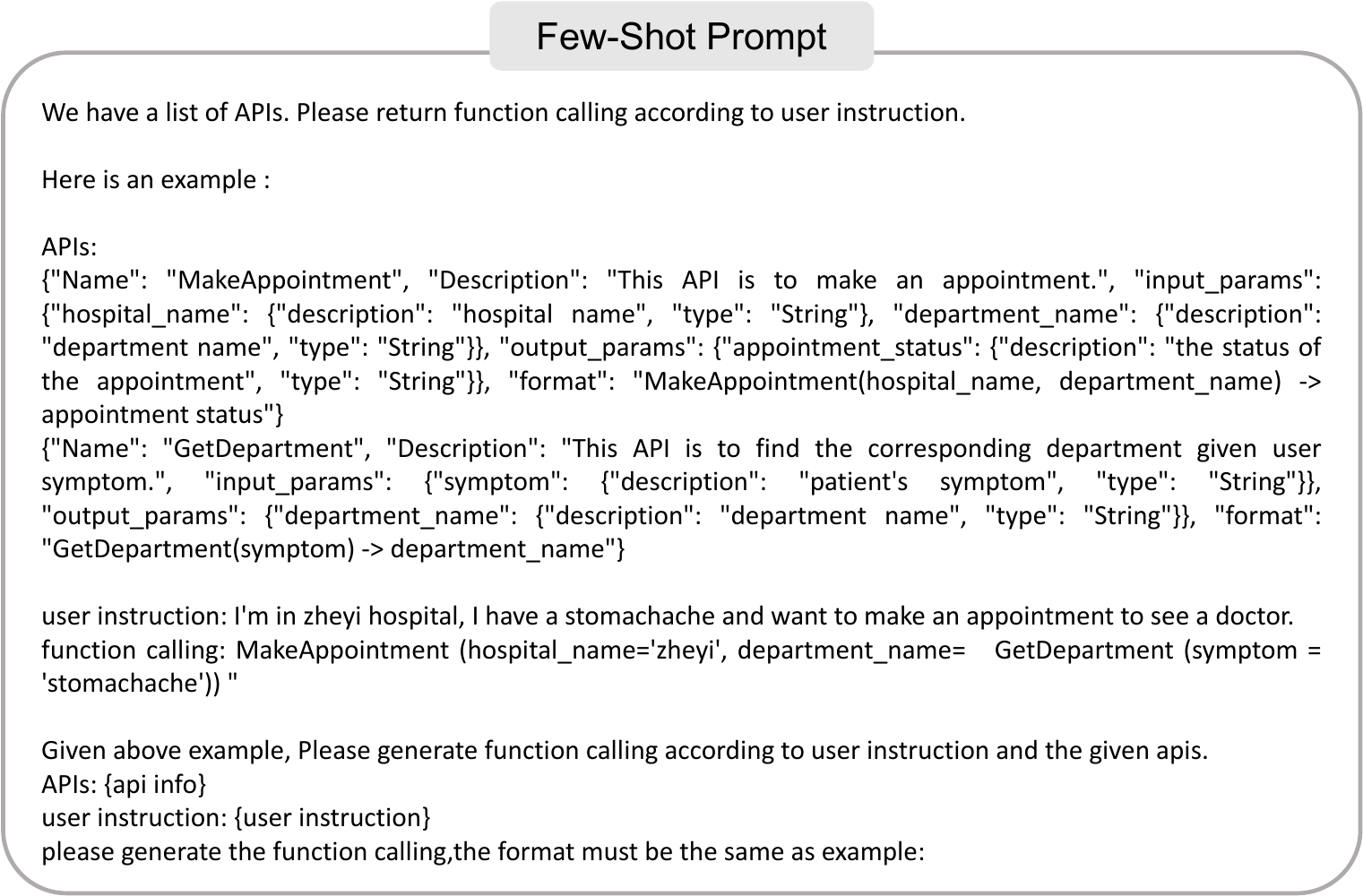}
  \caption{Prompt for Few-Shot method.} \label{Few-Shot}
\end{figure*}

\begin{figure*}[h]
  \centering
  \includegraphics[width=1 \linewidth]{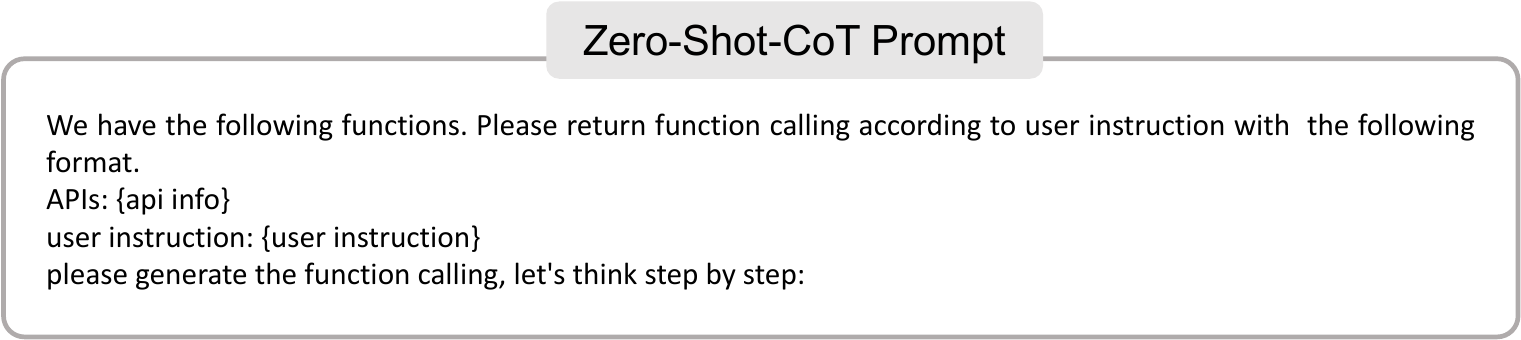}
  \caption{Prompt for Zero-Shot-CoT method.} \label{Zero-Shot-CoT}
\end{figure*}

\begin{figure*}[h]
  \centering
  \includegraphics[width=1 \linewidth]{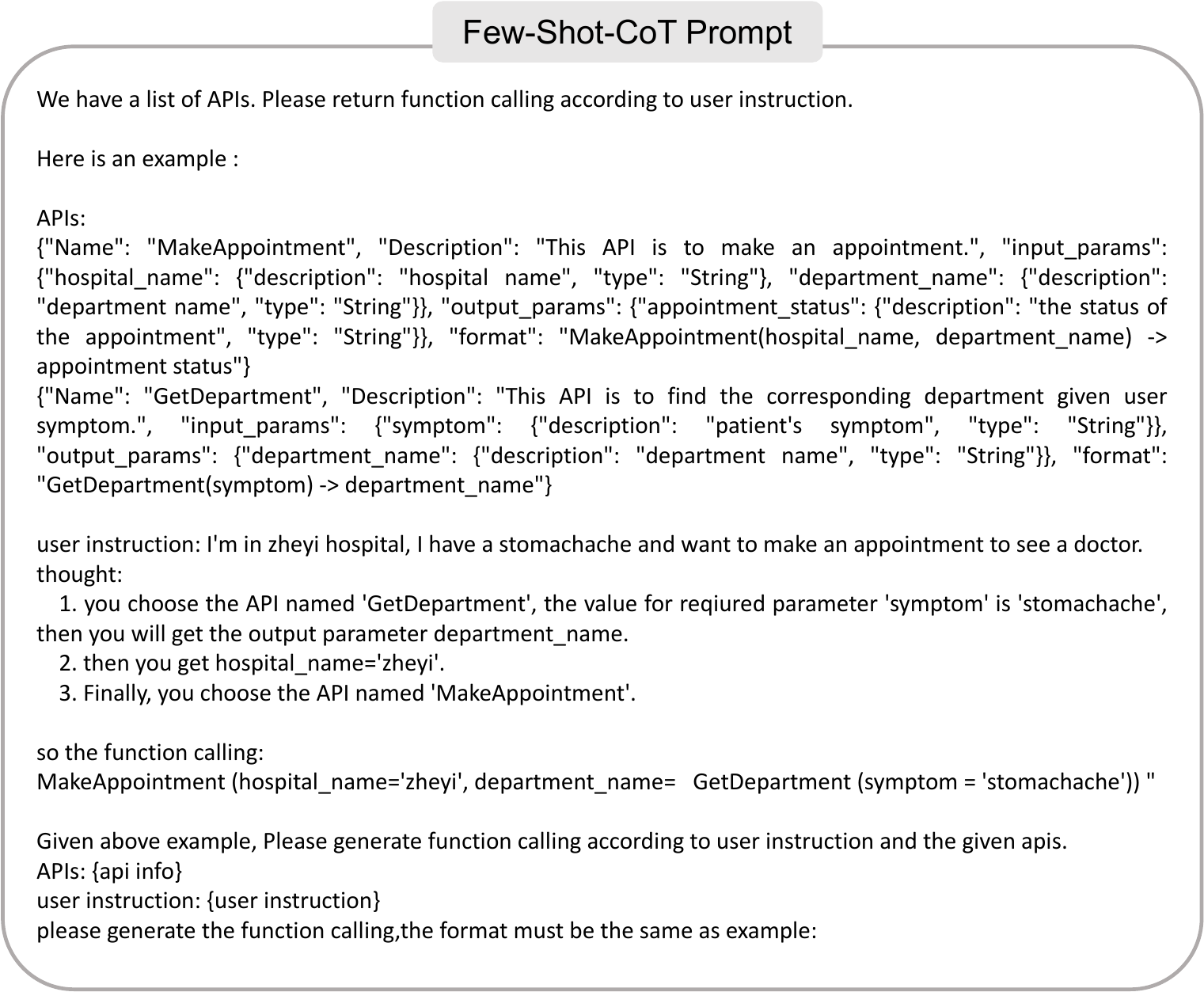}
  \caption{Prompt for Few-Shot-CoT method.} \label{Few-Shot-CoT}
\end{figure*}

The prompt for baseline methods are listed in Figure \ref{Zero-Shot}, Figure \ref{Few-Shot}, Figure \ref{Zero-Shot-CoT} and Figure \ref{Few-Shot-CoT}.

\subsection{Prompts for evaluation}\label{evaluation promot}
\begin{figure*}[h]
  \centering
  \includegraphics[width=1 \linewidth]{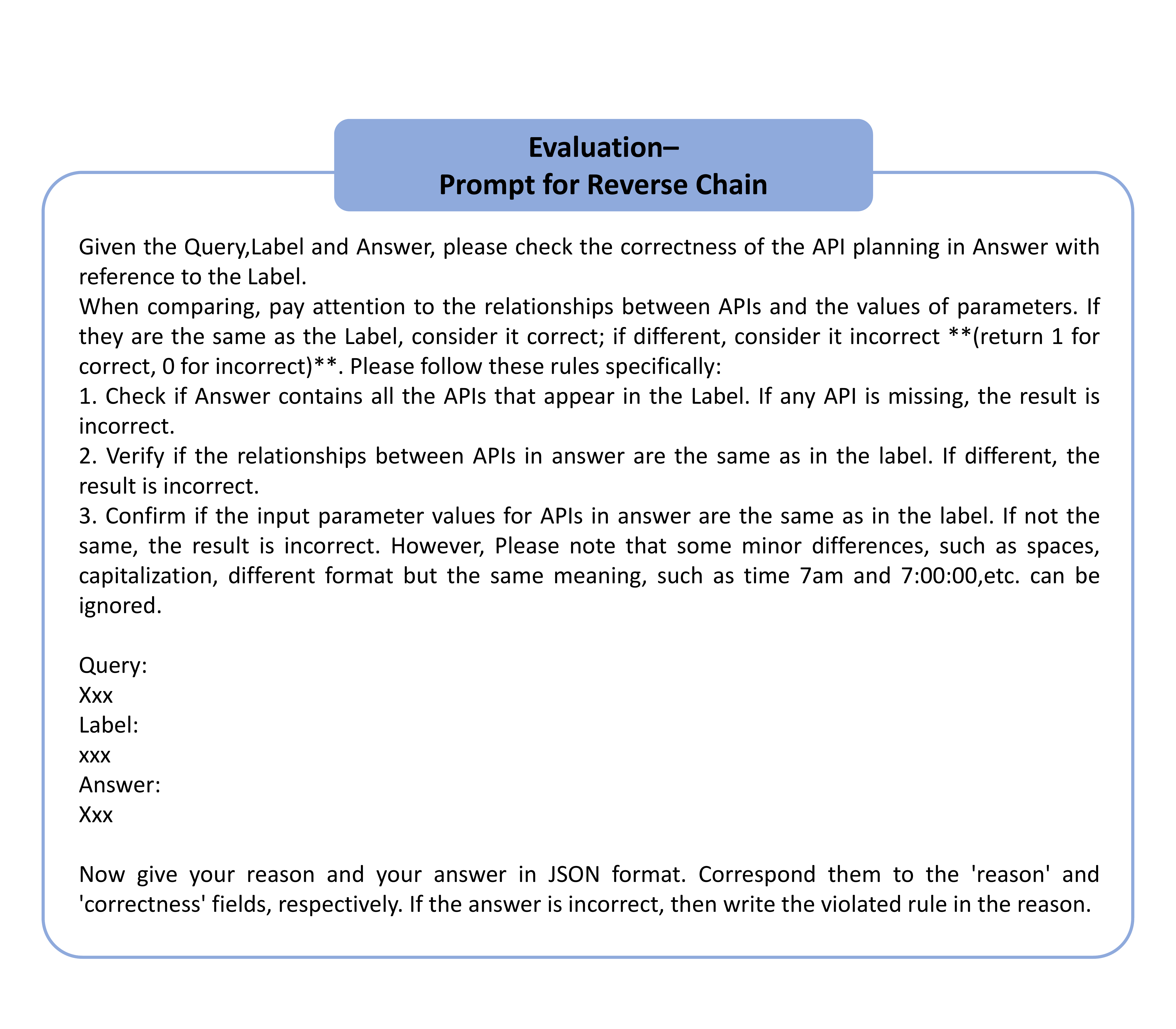}
  \caption{Prompt for evaluation for Reverse Chain.} \label{eval_rc}
\end{figure*}

\begin{figure*}[h]
  \centering
  \includegraphics[width=1 \linewidth]{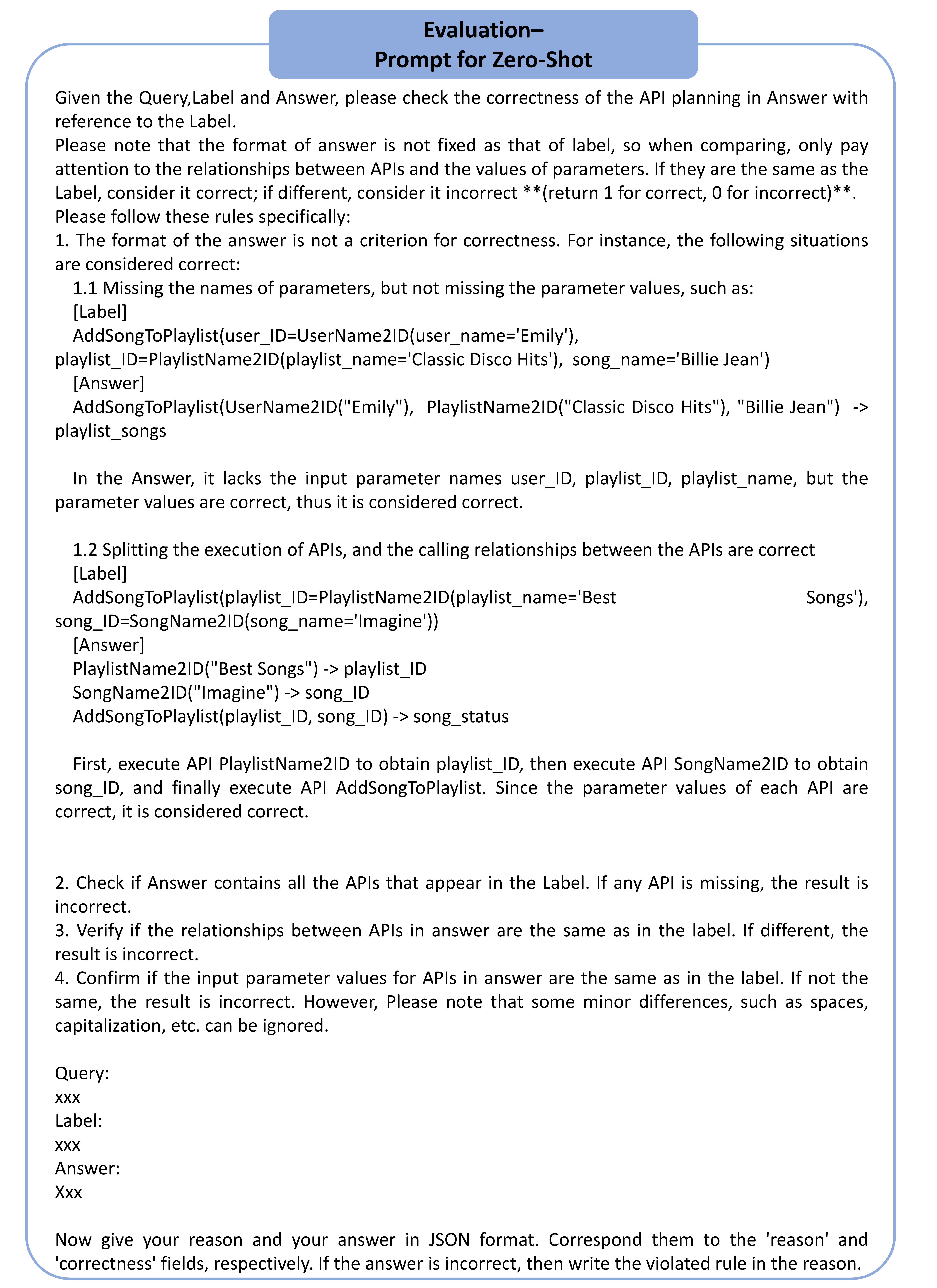}
  \caption{Prompt for evaluation for Zero-Shot.} \label{eval_zero_shot}
\end{figure*}

\begin{figure*}[h]
  \centering
  \includegraphics[width=1 \linewidth]{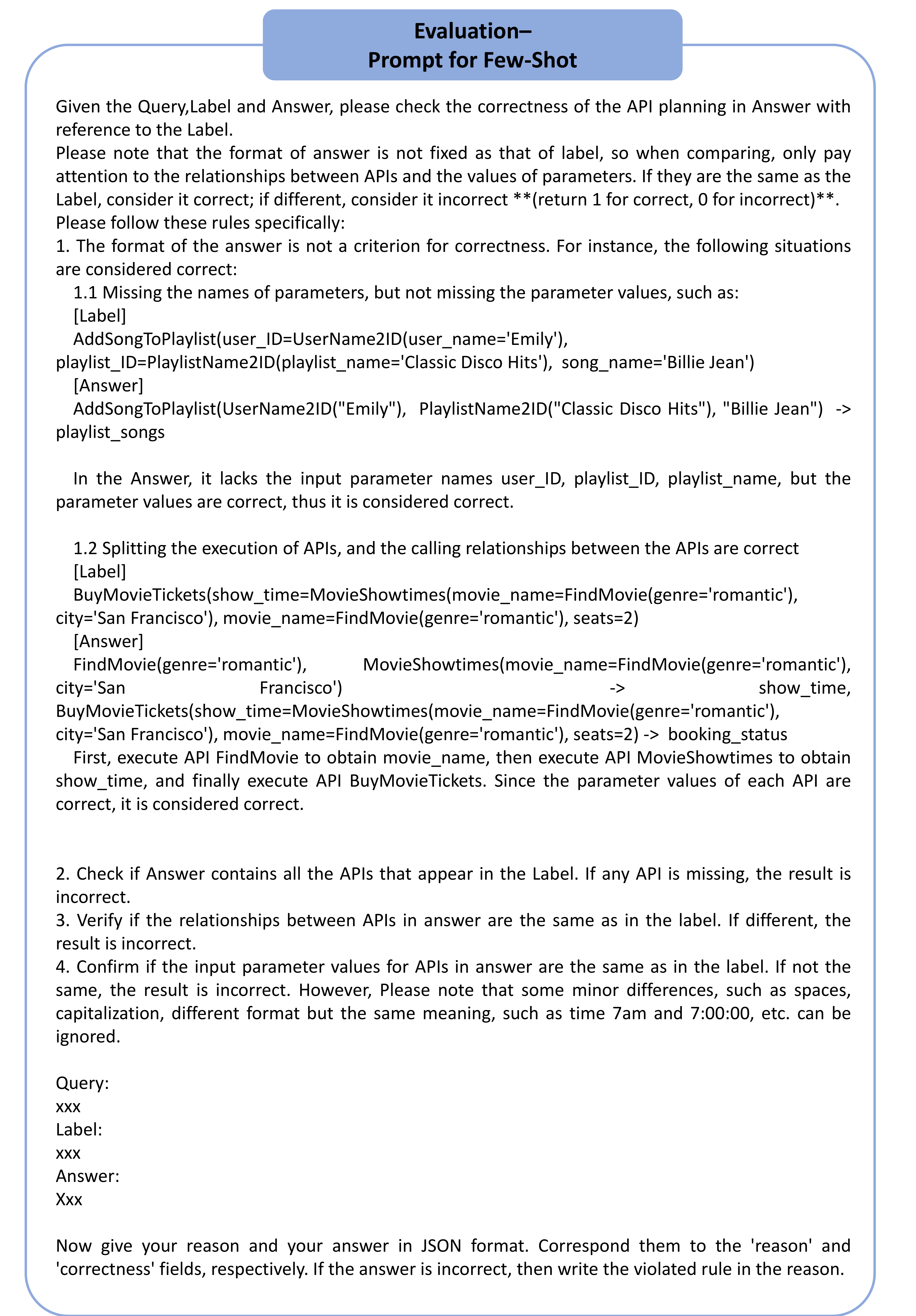}
  \caption{Prompt for evaluation for Few-Shot.} \label{eval_few_shot}
\end{figure*}

\begin{figure*}[h]
  \centering
  \includegraphics[width=1 \linewidth]{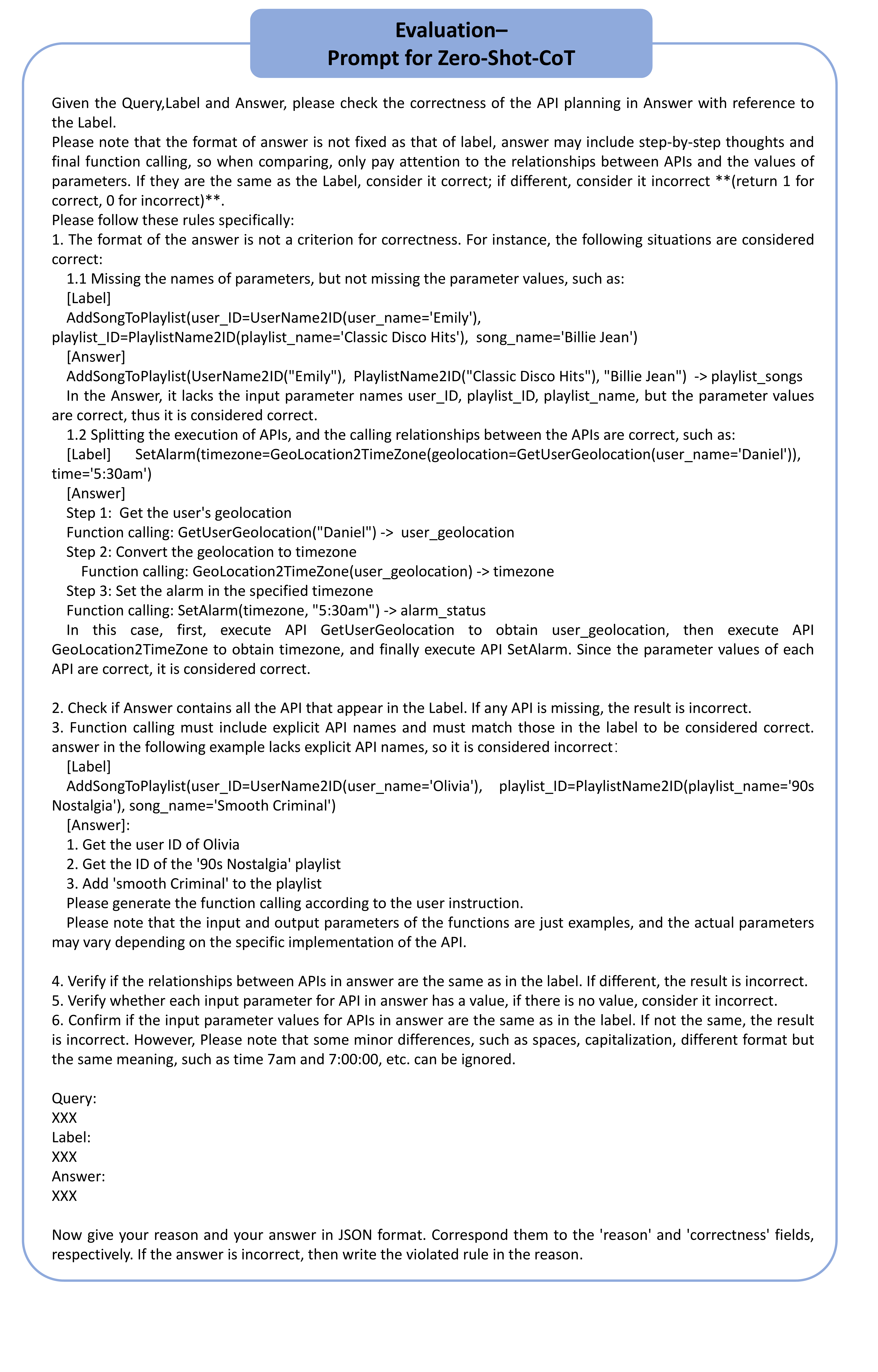}
  \caption{Prompt for evaluation for Zero-Shot-CoT.} \label{eval_zero_shot_cot}
\end{figure*}

\begin{figure*}[h]
  \centering
  \includegraphics[width=1 \linewidth]{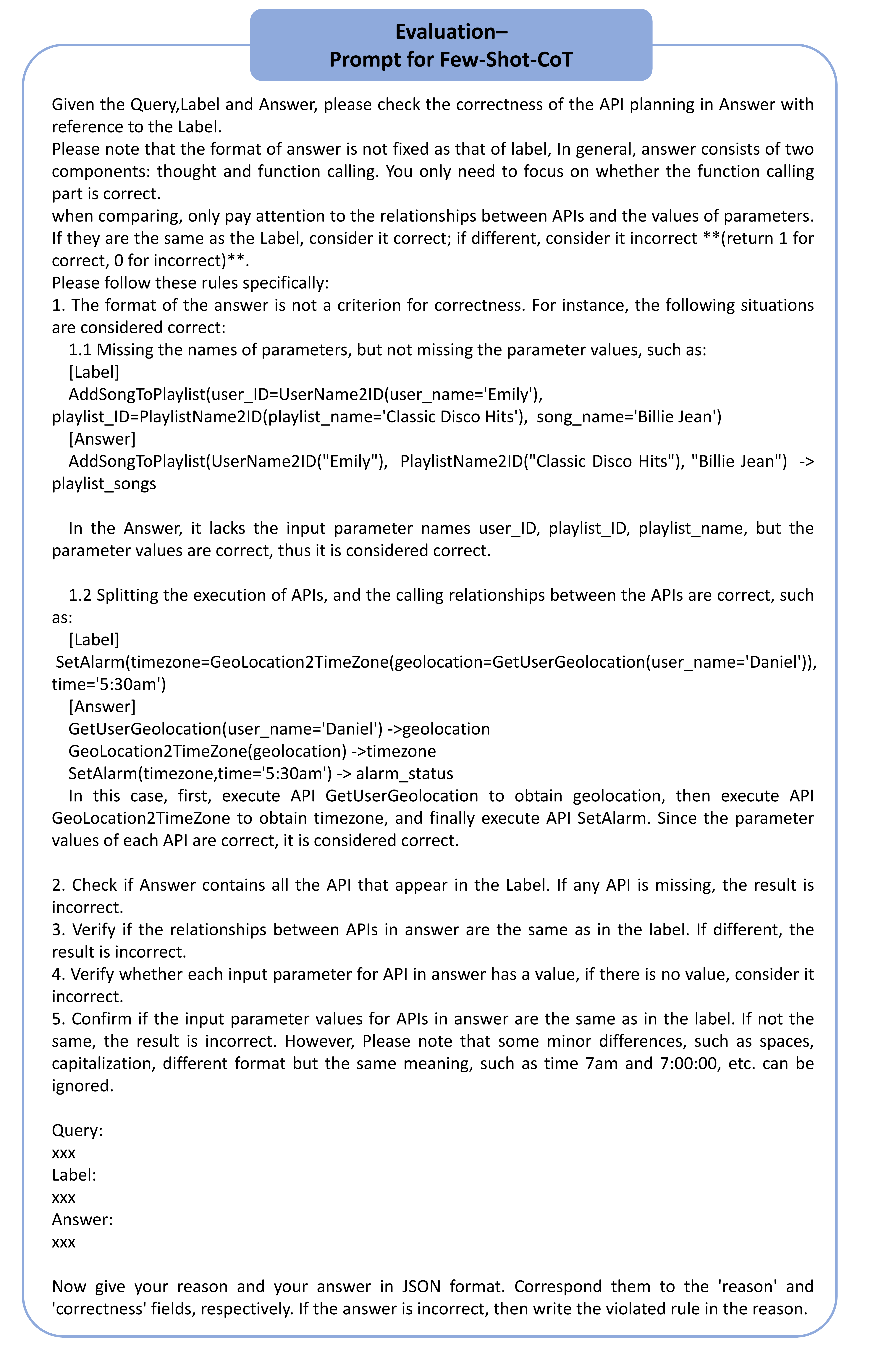}
  \caption{Prompt for evaluation for Few-Shot-CoT.} \label{eval_few_shot_cot}
\end{figure*}

\begin{figure*}[h]
  \centering
  \includegraphics[width=1 \linewidth]{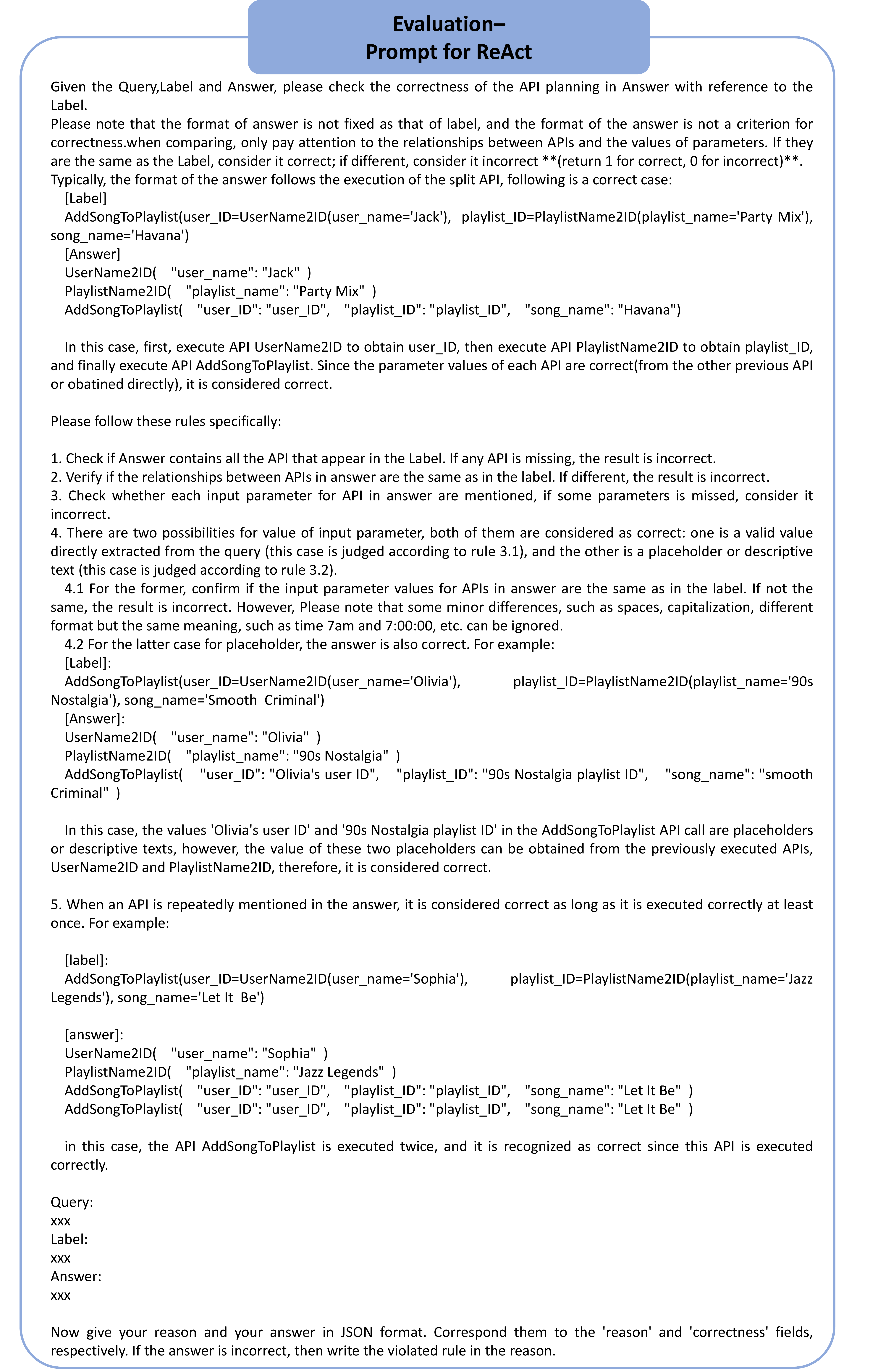}
  \caption{Prompt for evaluation for ReAct.} \label{eval_react}
\end{figure*}

Following the evaluation method used by \cite{tang2023toolalpaca}, We use GPT-4 as our evaluator. The evaluation prompts for different methods are shown in Figure \ref{eval_rc}, \ref{eval_zero_shot}, \ref{eval_few_shot}, \ref{eval_zero_shot_cot},\ref{eval_few_shot_cot},\ref{eval_react}. It should be noted that prior to conducting the ReAct evaluation, it is necessary to preprocess the answer to extract the function callings.

\end{document}